# Electrically Tunable Spin Polarization in a Carbon-Nanotube Spin Diode

Christopher A. Merchant and Nina Marković

*Department of Physics and Astronomy, Johns Hopkins University*

*Baltimore, MD 21218*

We have studied the current through a carbon nanotube quantum dot with one ferromagnetic and one normal-metal lead. For the values of gate voltage at which the normal lead is resonant with the single available non-degenerate energy level on the dot, we observe a pronounced decrease in the current for one bias direction. We show that this rectification is spin-dependent, and that it stems from the interplay between the spin accumulation and the Coulomb blockade on the quantum dot. Our results imply that the current is spin-polarized for one direction of the bias, and that the degree of spin polarization is fully and precisely tunable using the gate and bias voltages. As the operation of this spin diode does not require high magnetic fields or optics, it could be used as a building block for electrically controlled spintronic devices.

The ability to create, manipulate and detect spin currents is central to the development of spintronic devices [1]. Controlling the spin currents by purely electrical means [2, 3] is particularly interesting, as that would allow the integration of spintronic devices with conventional electronics. Creating spin-polarized currents typically involves injection from ferromagnets [4, 5] or magnetic semiconductors [6, 7]. Spin filters, spin memory devices and spin pumps using quantum dots with normal metal leads have also been proposed [8, 9]. If a quantum dot is connected to two ferromagnetic leads with different polarizations, it can operate as a spin-diode, as predicted recently [8, 10 - 14]. In this system, the spin-dependent asymmetry in the tunneling rates between the two junctions is predicted to lead to spin accumulation, current rectification and spin-polarized currents. If a quantum dot is attached to one ferromagnetic and one normal lead, the effect is even more pronounced [15 - 17] and can potentially be used for spin injection [18]. The current polarization in such a system can be controlled entirely by the gate and bias voltages. In this Letter, we report a pronounced asymmetry in the current-voltage characteristics of a quantum dot with one ferromagnetic and one normal metal lead. The observed asymmetry is analyzed in terms of spin-dependent tunneling rates and the results show excellent agreement with the theoretical predictions. Our device therefore represents an experimental realization of a tunable spin diode.

The sample consists of a carbon nanotube quantum dot connected to cobalt and niobium electrodes. Carbon nanotubes (CNTs) [19, 20] are ideal candidates for fabricating spin-based devices because of their long spin-coherence lengths [21]. CNTs also typically form tunnel barriers with metallic contacts so that small nanotube sections contacted by metal electrodes behave like quantum dots at low temperatures [22]. An atomic force microscope image of the sample is shown in Fig. 1(a). To create the sample, a niobium lead was first defined using electron-beam lithography and deposited by sputtering. Single-walled carbon nanotubes were then grown on powder catalyst in flowing methane using the chemical vapor deposition technique [23]. As-grown nanotubes were dispersed in 1,2-dichloroethane and sonicated to create a CNT suspension. A single carbon nanotube was positioned on top of the niobium lead using ac-dielectrophoresis of the CNT suspension [24]. Finally, the cobalt lead was thermally evaporated on top of the carbon nanotube after another electron-beam lithography step. Measurements were performed by applying a bias voltage and measuring the current through the sample. A capacitively-coupled gate voltage was applied to the sample through a thermally-grown $SiO_2$ layer, as shown in the measurement schematic in Fig. 1(b). All measurements were carried out at temperatures of 10K, above the superconducting transition temperature of niobium. We have studied several samples, with the sections of carbon nanotubes between the leads typically between 250-500 nm long. In this work, we present the results obtained from the sample shown in Fig. 1(a).

A map of the two-terminal differential conductance, $G=dI/dV$, is shown as a function of bias and gate voltages in Fig. 2(a). We observe typical quantum dot behavior, characterized by the diamond-shaped regions where the electron transport through the device is forbidden due to the Coulomb blockade [25]. From the sizes of the Coulomb diamonds we determine the charging energy and level spacing of the dot to be 7 meV and 3 meV, respectively [26]. For the range of gate and bias voltages shown in Fig. 2(a), the Coulomb diamonds are regular and do not vary in size. In this region, we do not observe the even-odd filling pattern of the quantum dot, typically seen when the energy levels on the dot are spin-degenerate [27]. The Coulomb blockade is overcome at the edges of the diamonds in Fig. 2(a) as the Fermi level of one of the leads becomes resonant with the ground-state level of the dot. We find that only the ground-state level of the dot is reached for a large range of bias voltages due to the large energy level spacing of the dot.

The conductance as a function of gate voltage for positive ($G_+$) and negative ($G_-$) bias is shown in Fig. 2(b). Conductance data are taken at bias voltages of ± 5mV as indicated by dashed lines in Fig. 2(a). When the device is negatively biased, the electrons tunnel out of the ferromagnetic lead, onto the dot, and into the normal lead. The data have been fit using the Breit-Wigner lineshape model [25], using the capacitance ratio $\alpha = 0.03$ [22] and a temperature of 10K. The locations of dot energy levels in the gate voltage, as well as maximum conductance levels, are the fitting parameters of this model. A pronounced asymmetry between the positive-bias and negative-bias conductance traces is observed in Fig. 2(b). In particular, around the gate voltage values of $V_g$ = 8.45V, 8.85V and 9.25V, the peaks in $G_+$ are strongly suppressed relative to the corresponding peaks in $G_-$. The normal-metal lead is resonant for these values of gate voltage. Defining an average difference in conductance as $\Delta G = (G_- - G_+)/G_-$, we measure $\Delta G$ to be 16 ± 2% for ferromagnetic resonance peaks compared to a 38 ± 6% difference when the normal-lead lead is resonant.

An asymmetry in conductance with respect to bias direction is typically observed in quantum dots which do not have symmetrically coupled leads. Since our leads are made of two different materials (cobalt and niobium), it is reasonable to expect that their coupling strengths to the carbon nanotube dot are not identical. However, as we will show below, the observed asymmetry cannot be explained entirely by asymmetric coupling to the leads. We will argue that there is a significant contribution stemming from the ferromagnetic polarization of one of the leads, which leads to a spin-dependent suppression of tunneling when electrons are coming from the resonant normal lead.

To highlight this unusual contribution to the observed asymmetry, we analyze the single-junction conductances for each of the leads. The single-junction conductances are labeled as $G_{N(FM)}$, where N and FM denote the normal metal and the ferromagnet, respectively. An additional subscript (+ or –) refers to either positive or negative bias. The single-junction conductances were not measured directly, but were obtained as a part of the Breit-Wigner fitting procedure for the data shown in Fig. 2(b). Since a normal tunnel junction has to be symmetric with respect to bias direction [26], we correct for the asymmetry caused by the quantum dot-electrode coupling by scaling the single-junction conductances so that $G_{N+}$ and $G_{N-}$ match. Observing the scaled single-junction conductances in Fig. 3, we find that the asymmetry with respect to the bias direction is qualitatively different for the normal and the ferromagnetic junction. Both $G_{N+}$ and $G_{FM+}$ have been scaled by a factor of 1.15, however only the normal junction conductances match, as shown in Fig. 3(a). The scaled ferromagnet-dot junction conductance still displays an asymmetry, which is apparent in Fig. 3(b). In particular, $G_{FM+}$ is lower than $G_{FM-}$ only for certain values of gate voltage, which means that the tunneling is suppressed in the case of positive bias as compared to the negative bias. This can only occur if the electrons tunnel more easily from the ferromagnetic lead onto the dot than vice versa for those particular values of gate voltages.

To describe the nature of this asymmetry in a more quantitative way, we assign different tunneling rates to the two tunnel junctions: $\Gamma_{FM}$ refers to the tunneling rate through the ferromagnetic contact, while $\Gamma_N$ refers to the tunneling rate through the normal contact. As the gate voltage is varied, the ground-state level on the dot moves in and out of resonance with the two leads, modifying the ratio of the two tunneling rates and creating peaks and valleys in the conductance. Both tunneling rates depend on the available energy levels on the dot, and it is reasonable to assume that $\Gamma_{FM}$ also depends on the spin polarization of the ferromagnetic electrode. For the ferromagnetic lead we take $\Gamma_{FM} = \Gamma_0(1\pm p)$ [11], where p is the thermodynamic polarization of the ferromagnet, positive (negative) sign is for majority (minority) spin, and $\Gamma_0$ is a constant.. The normal-metal lead is not spin-dependent, so we take the tunneling rate to be $\Gamma_N = \gamma\Gamma_0$ [11]. The factor $\gamma$ is the ratio of the single-junction conductances ($\gamma=G_N/G_{FM}$), which accounts for the asymmetry in the electrode couplings as a function of the gate voltage. As we will show, the dependence of $\Gamma_{FM}$ on the spin polarization of the ferromagnetic lead can be detected through the asymmetry of the current with respect to the direction of the bias. In particular, the fact that $\Gamma_{FM}$ is different for the majority and minority spins leads to a spin-dependent current rectification.

We start with an expression for the current [15] derived by employing the master equation approach [29] in the sequential tunneling regime. Using the tunneling rates defined above, we find the total current through a quantum dot with a single ferromagnetic lead to be

$$I = \begin{cases} 2\Gamma_0 e\gamma \dfrac{1-p^2}{(1-p^2)+2\gamma} & eV > 0 \\ -2\Gamma_0 e\gamma \dfrac{1}{2+\gamma} & eV < 0 \end{cases} \quad (1)$$

for the case of positive and negative bias, respectively. In this expression, the current depends on the gate voltage through the factor γ, and can be directly compared to the experimental data. Fig. 4(a) shows that the negative-bias current calculated from Eq. (1) is in excellent agreement with the measured negative-bias current as a function of gate voltage. The positive-bias current is analyzed in Fig. 4(b) for three different values of polarization p: 0.0, 0.3 and 0.4. We find the calculated current for the p=0.0 case does not fit the data, while there is a good agreement when the current is calculated with a p value between 0.3 and 0.4. This result agrees with the expected thermodynamic polarization value for cobalt, p = 0.4 [28].

We further analyze the deviation of the positive-bias current from the p=0.0 curve by defining the current difference, ΔI, as ΔI = ($I_0$ - $I_+$)/($I_0$), where $I_0$ is the calculated current for the p=0.0 case and $I_+$ is the measured current for positive bias. This current difference depends strongly on the gate voltage, with peaks ranging from 8% to 17%, as shown in Fig. 4(c). It should be emphasized that ΔI shown in Fig. 4(c) describes only the deviation from the p=0.0 line, and is only observed in the case of positive bias.

It is important to note that the current difference ΔI peaks for the values of gate voltage at which the normal lead is resonant with the single energy level on the dot. In this situation, the conductance is dominated by the non-resonant tunnel junction between the ferromagnetic electrode and the quantum dot. The tunneling rates for the majority and minority spins are different ($\Gamma_{FM}^\uparrow \neq \Gamma_{FM}^\downarrow$) for this junction, meaning that there are fewer available states for the minority spins in the ferromagnetic electrode. As a result, the minority spins will spend more time on the dot, leading to spin accumulation [15]. This leads to a decreased current due to electrons with minority spins, resulting in a decreased total conductance through the device for positive bias. This spin-diode effect depends on the fact that there is a single non-degenerate level available on the dot in the bias window. As described theoretically in detail in Refs. 11 and 15, our results imply that the current through the device is spin-polarized for specific values of gate and bias voltages. When the gate and bias voltage are tuned away from this regime, the current rectification is not observed, and the current is not spin-polarized. In particular, when the bias is increased to allow transport through another energy level on the dot, the spin diode effect is diminished or destroyed by spin flips.

In summary, we have observed a clear decrease in the current through a quantum dot with one ferromagnetic and one normal metal lead for one direction of bias voltage. Our analysis shows that this rectification occurs because of the asymmetric tunneling rates between the ferromagnetic lead and the dot for the majority and minority spins. The observed asymmetry implies the existence of spin-polarized currents, which are fully tunable by a proper choice of gate and bias voltages. By increasing the polarization of the ferromagnet, the diode effect could be increased up to a $\Delta I$ value of 100% [30]. Such carbon-nanotube spin diodes could easily be positioned using ac-dielectrophoresis [24], and could be operated at room temperature by reducing their size [31]. Additionally, with the proper choice of the material for the non-magnetic electrode [32], this device could be invaluable for achieving controllable spin injection.

We thank J.C. Egues for useful comments. This work was supported in part by the National Science Foundation under grants DMR-0547834 and DMR-0520491 (MRSEC), Alfred P. Sloan Foundation under grant BR-4380, and ACS PRF # 42952-G10.

FIG. 1. (a) Atomic force microscope image of the sample (scale bar: 1μm). The distance between the electrodes is 300nm. CNT height is measured as < 2 nm. The electrodes are cobalt and niobium, as indicated on the image. (b) Schematic of the measurement setup. Bias is applied between the niobium and the cobalt lead with niobium lead grounded. Gate voltage is applied to the doped silicon substrate through a 500nm thick thermally grown layer of $SiO_2$.

FIG. 2. (a) Differential conductance, $G=dI/dV$, for varying gate and bias voltages with regular coulomb diamonds. Black represents zero conductance and yellow indicates a maximum conductance of 0.06 $e^2/h$. Dashed lines indicate the bias voltage of 5mV. (b) Conductance as a function of gate voltage at 5 mV. Open blue circles indicate negative bias while full green triangles represent positive bias. Positive bias data have been shifted in the gate voltage by $\Delta V_G = 10$ mV/$\alpha = 0.34$ V to allow direct comparison between corresponding conductance peaks. Theoretical fits (represented by solid lines) come from the Breit-Wigner model with locations of conductance peaks as a fitting parameter.

FIG. 3. (a) Normal-lead conductance for positive (full diamonds) and negative (open diamonds) bias. Positive bias data have been scaled by a factor of 1.15 so that $G_{N+} = G_{N-}$. (b) Ferromagnetic-lead conductance for positive (full circles) and negative (open circles) bias. Positive bias data ($G_{FM+}$) have been scaled by the same factor as $G_{N+}$.

FIG. 4. (a) Negative-bias current as a function of gate voltage. Open black squares represent measured current. Solid line is a fit obtained from Eq. (1) for eV < 0 case. (b) Positive-bias current as a function of gate voltage. Open black squares represent measured current. Theoretical fits to the data use eV > 0 case of Eq. (1) with p=0.0 (dotted red), 0.3 (dashed blue), & 0.4 (solid green). Conductance ratio, γ, in Eq. (1) is taken from the negative bias data in Fig. 3. (b) ΔI as a function of gate voltage for a fixed bias voltage (5 mV). ΔI calculated as difference from theoretical p=0.0 fit in Fig. 4b.

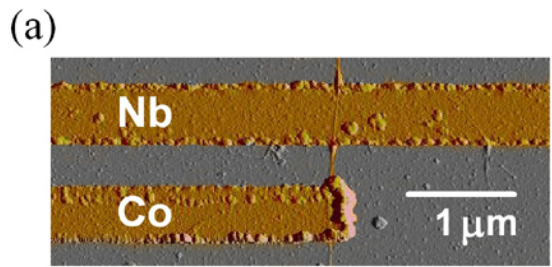

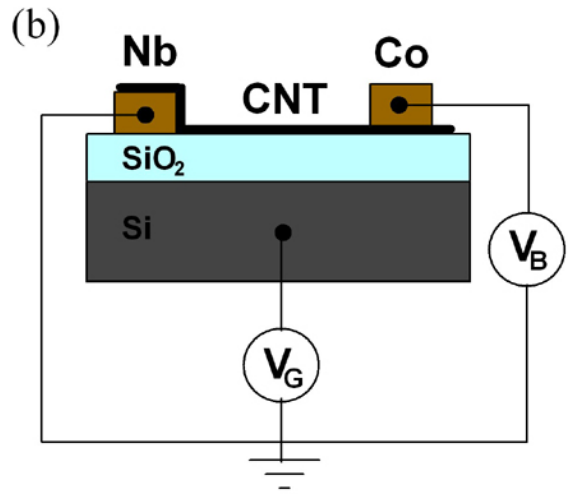

Fig. 1

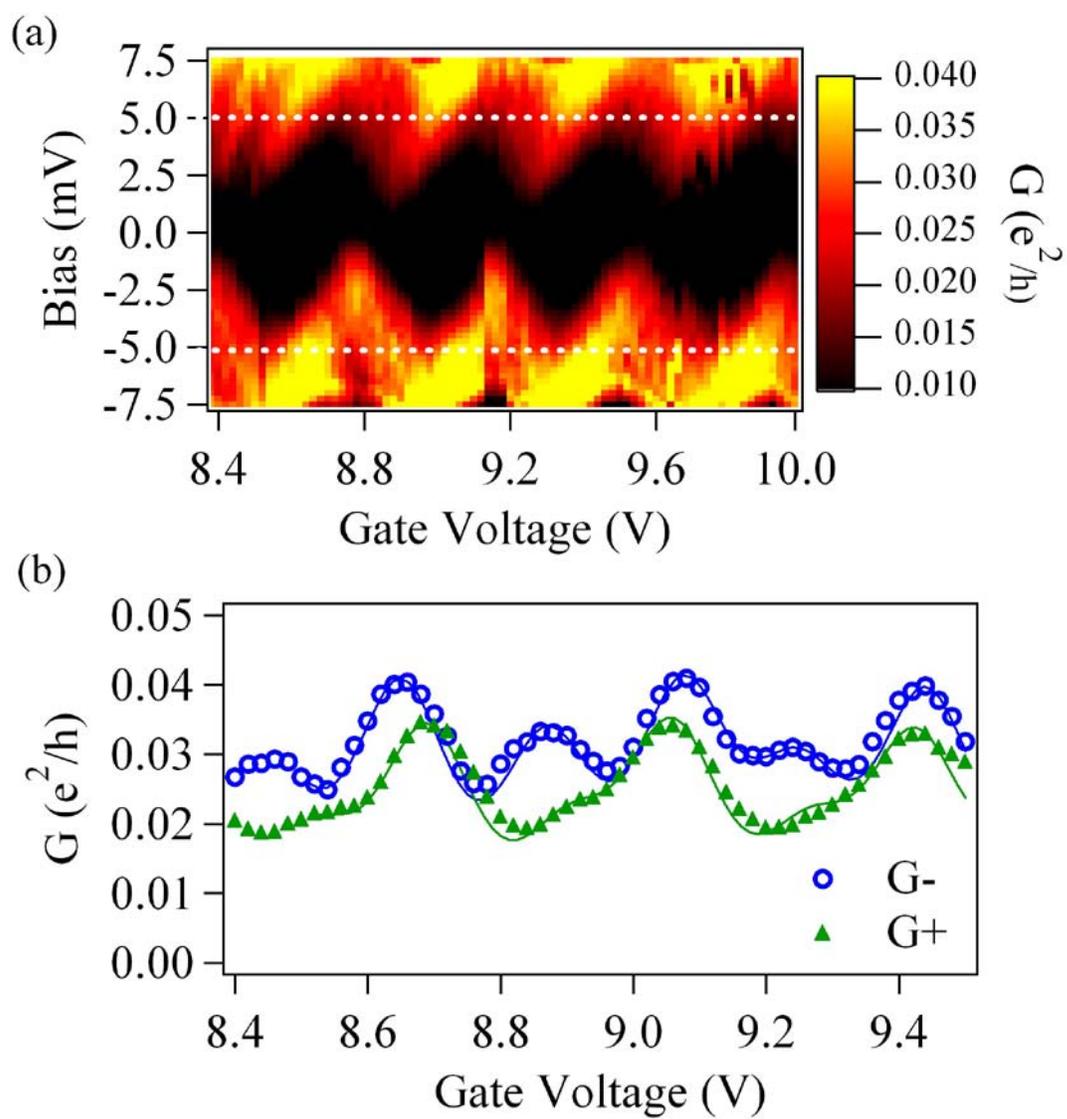

Fig. 2

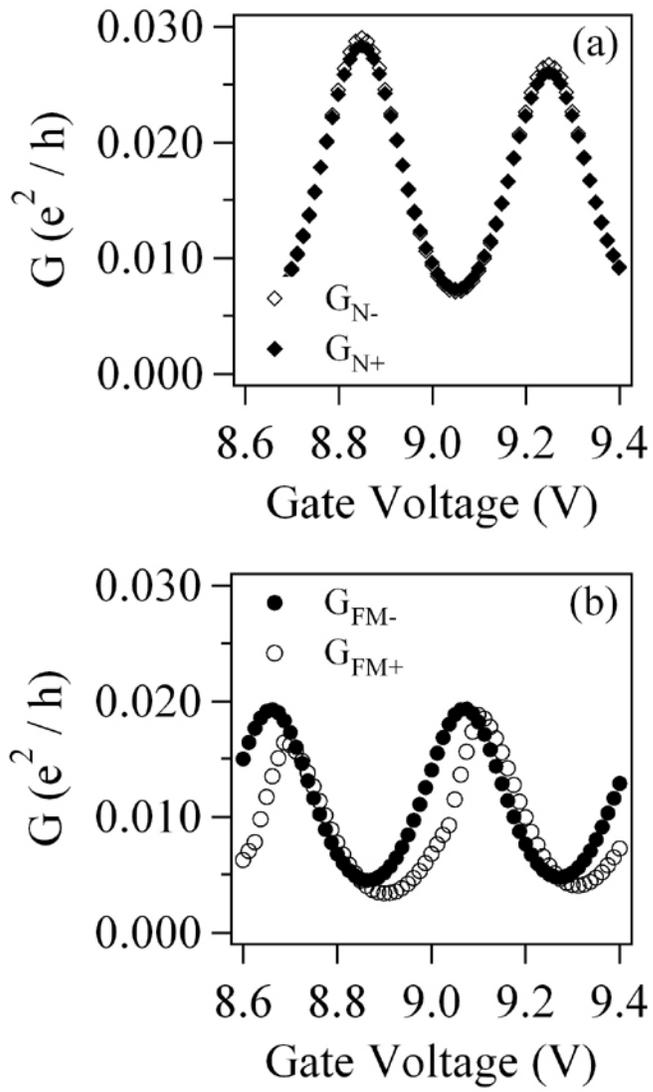

Fig. 3

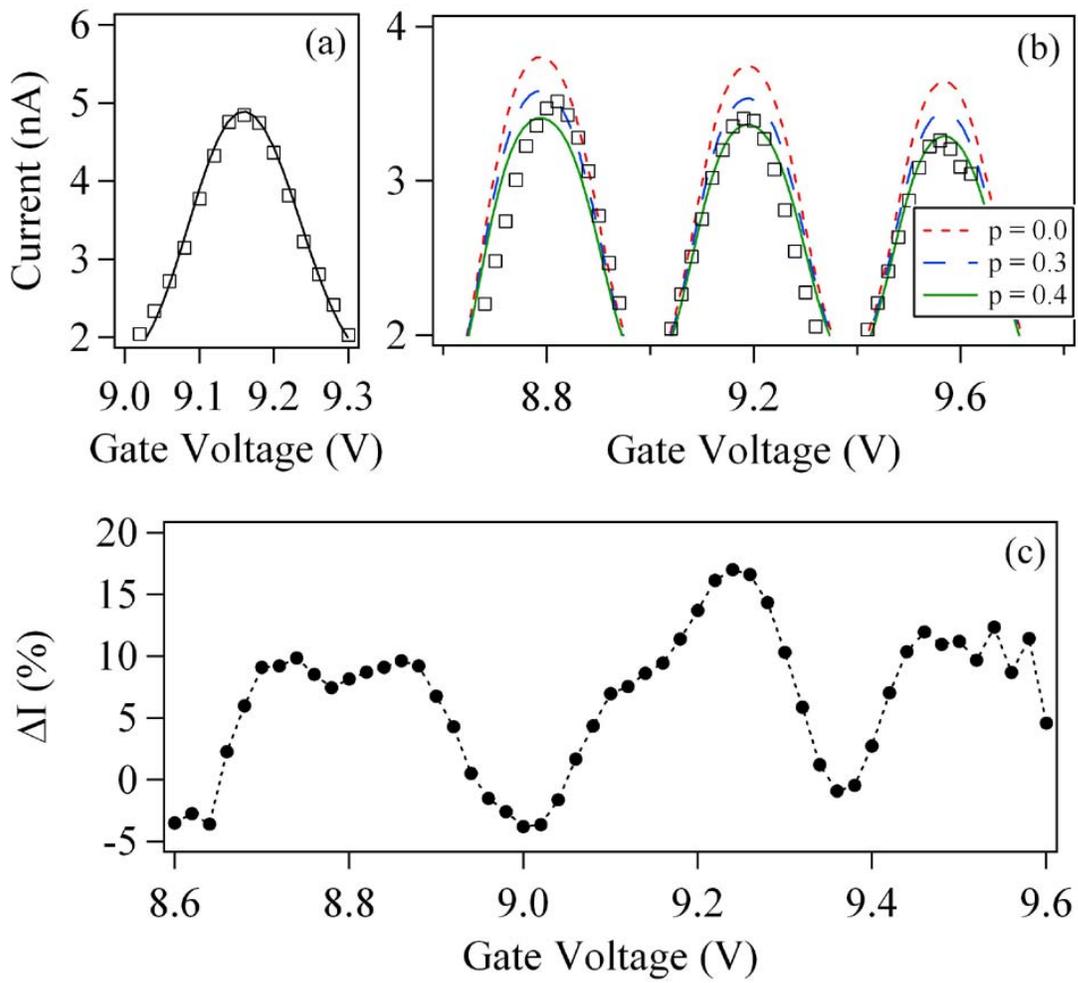

Fig. 4